\documentclass{ecai}
\usepackage{graphicx}
\usepackage{latexsym}
\usepackage{float}
\usepackage{makecell}
\usepackage{colortbl}  
\usepackage{xcolor}
\usepackage{cite}
\usepackage{algorithmic}
\usepackage{graphicx}
\usepackage{color}
\usepackage{subfig,graphicx}
\usepackage[justification=justified]{caption}
\usepackage[bottom]{footmisc}
\usepackage{graphicx}

\DefineFNsymbols*{newfootnote}{%
    \textasteriskcentered\dag\textdaggerdbl{\ding{73}}\P{**}%
    {\ding{172}}{\ding{173}}{\ding{174}}}
\setfnsymbol{newfootnote}

\begin{document}

\begin{frontmatter}

\title{ Instance-Wise Adaptive Tuning  and Caching for Vision-Language Models}

\author{\fnms{Chunjin}~\snm{Yang}}
\author{\fnms{Fanman}~\snm{Meng}\thanks{Corresponding Author. Email: fmmeng@uestc.edu.cn.}}
\author{\fnms{Shuai}~\snm{Chen}} 
\author{\fnms{Mingyu}~\snm{Liu}}
\author{\fnms{Runtong}~\snm{Zhang}}

\address{University of Electronic Science and Technology of China}

\begin{abstract}
Large-scale vision-language models (LVLMs) pre-trained on massive image-text pairs have achieved remarkable success in visual representations. However, existing paradigms to transfer LVLMs to downstream tasks encounter two primary challenges. Firstly, the text features remain fixed after being calculated and cannot be adjusted according to image features, which decreases the model's adaptability. Secondly, the model’s output solely depends on the similarity between the text and image features, leading to excessive reliance on LVLMs. To address these two challenges, we introduce a novel two-branch model named the Instance-Wise \textbf{A}daptive \textbf{T}uning  and \textbf{C}aching (\itshape ATC\rm). Specifically, one branch implements our proposed $ConditionNet$, which guides image features to form an adaptive textual cache that adjusts based on image features, achieving instance-wise inference and improving the model's adaptability. The other branch introduces the similarities between images and incorporates a learnable visual cache, designed to decouple new and previous knowledge, allowing the model to acquire new knowledge while preserving prior knowledge. The model's output is jointly determined by the two branches, thus overcoming the limitations of existing methods that rely solely on LVLMs. Additionally, our method requires  limited computing resources to tune parameters, yet outperforms existing methods on 11 benchmark datasets.
\end{abstract}

\end{frontmatter}

\section{INTRODUCTION}
%

Large-scale vision-language models (LVLMs) are trained through contrastive learning on a vast amount of image-text pairs. These models map images and texts to the same space through textual encoders and visual encoders. LVLMs, for instance CLIP \cite{radford2021learning}, ALIGN\cite{cohen1997align}, and  ALBEF\cite{li2021align}, have shown excellent  performance in downstream tasks such as semantic segmentation\cite{zhou2022zegclip,xu2023side}, object detection\cite{zareian2021open,ma2022open}, VQA\cite{song2022clip}, and so on. However, LVLMs have a large number of parameters, for example, the CLIP\cite{radford2021learning} model has 428 millions parameters, directly fine-tuning a model could potentially compromise the valuable knowledge obtained during the large-scale pre-training phase,  and can pose a risk of over-fitting to the downstream task.

There are currently two main paradigms, as shown in Figure \ref{exsting}, to address above issues: input-level prompt, such as CoOp\cite{zhou2022learning}, CoCoOp\cite{zhou2022conditional}, and feature-level adapter, such as CLIP-Adapter\cite{gao2021clip}, TaskRes\cite{yu2022task}. However, for input-level prompt, during the training process, damage would be susceptible to prior knowledge, as demonstrated by CoOp's\cite{zhou2022learning} 1-shot classification accuracy being lower than the Zero-shot CLIP\cite{radford2021learning}. For  feature-level adapter, due to the excessive coupling of prior and new knowledge, the models' ability to learn new knowledge is limited. Additionally, both paradigms suffer from the same problem, that is, the final classification result is only dependent on the similarity between textual features and visual features, leading to excessive reliance on LVLMs, and models' performance upper limit is determined by the LVLMs. 
Moreover, even within the same category of images, there are differences in visual features. Class-wise text features are not sufficient to cover the large changes in appearance context and geometry of the current category, making it impossible to adapt to the unique features of each test image. Therefore, if only fixed  class-wise text features are used as the final classification criteria, the stability of the model will be compromised. Neither paradigm provides a solution for this problem.
\begin{figure}[!t]
\centering
\includegraphics[width=8cm]{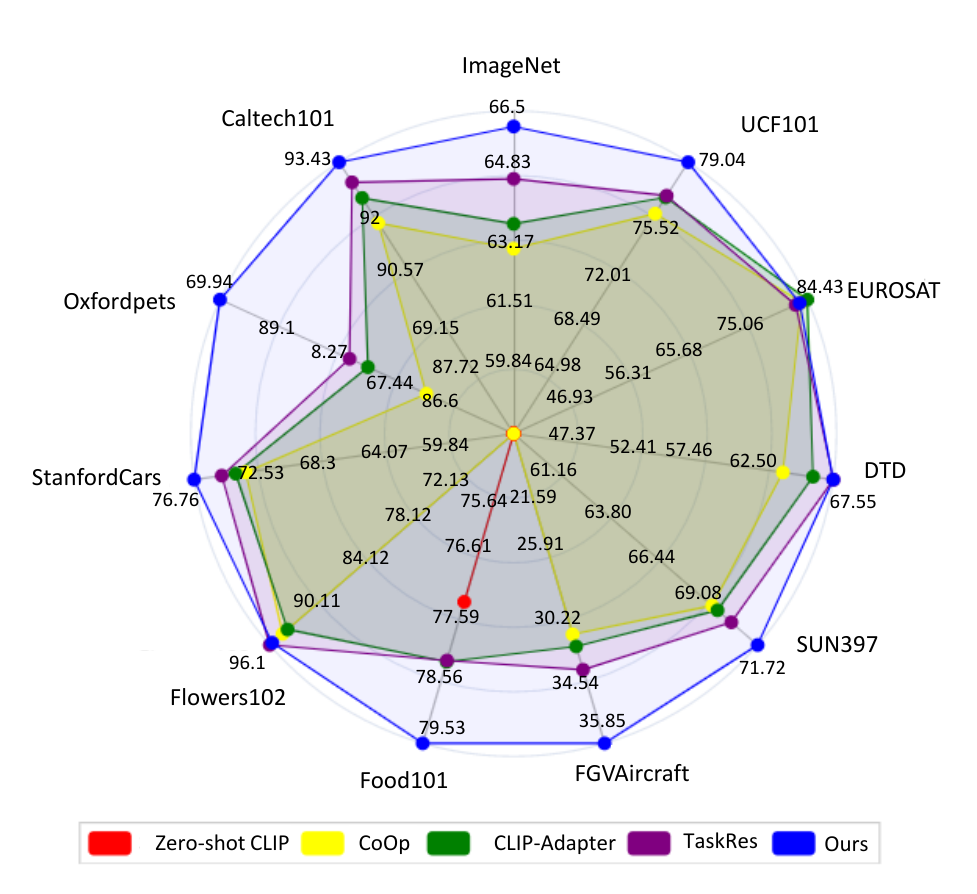}

\caption{Performance comparison between Zero-shot CLIP\cite{radford2021learning}, CoOp\cite{zhou2022learning}, CLIP-Adapter\cite{gao2021clip}, TaskRes\cite{yu2022task}  and our \itshape ATC \rm on ImageNet\cite{deng2009imagenet} with 16-shot settings.}

\label{few_shot_first}
\end{figure}

\begin{figure*}[!t]
    \begin{minipage}[t]{0.5\linewidth}
        \centering
        \includegraphics[width=\textwidth]{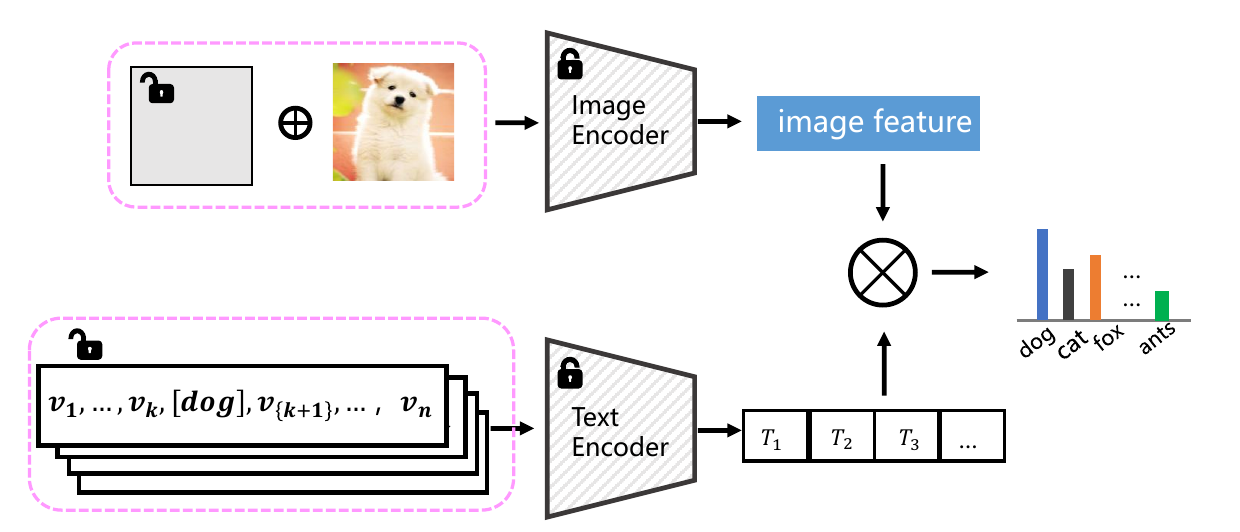}
        \centerline{(a)  Input-level prompt style}
    \end{minipage}%
    \begin{minipage}[t]{0.5\linewidth}
        \centering
        \includegraphics[width=\textwidth]{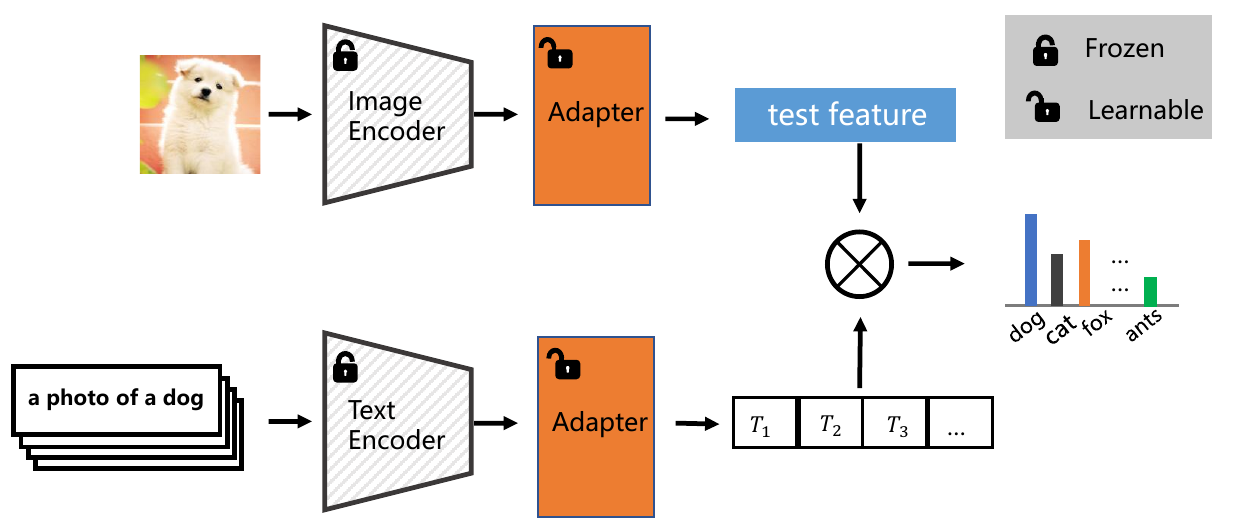}
        \centerline{(b) Feature-level adapter style}
    \end{minipage}
    \caption{\textbf{Exsiting paradigms}. (a)Input-level prompt style. Replace the originally discrete text or image  with continuous, learnable vectors. (b)Feature-level adapter style. Use an adapter after image or text features to fine-tune the features. }
    \label{exsting}
\end{figure*}

In response to the above-mentioned  challenges, we propose a novel two-branch model named \itshape ATC\rm: one branch introduces the similarities between train and val images, and employs training data to create a learnable visual cache, which decouples old and new knowledge and allows our method to retain previous knowledge of LVLMs while maximizing the acquisition of new knowledge. The other  branch uses our proposed $ConditionNet$ to direct the visual feature to generate  textual biases, which are overlaid on the text feature to produce an adaptive textual cache. This cache automatically adjusts the text feature based on the image characteristics, achieving instance-wise inference and enhancing the adaptability and generalization of the model. The final output of the model is a combination of the two branches, breaking the existing methods' excessive reliance on LVLMs. Additionally, our \itshape ATC \rm only requires minimal computing resources to be trained.

We benchmarked our \itshape ATC \rm on 11 datasets covering various visual recognition tasks such as classification for generic objects, scenes, actions, and fine-grained categories, and specialized tasks such as texture and satellite image recognition. Our results show that \itshape ATC \rm can effectively transfer pre-trained vision-language models to downstream tasks with limited data, and with better efficiency than existing methods, as demonstrated in Figure \ref{few_shot_first}. In summary, our contributions can be summarized as follows:

\begin{itemize}
\item We utilized our proposed $ConditionNet$ to guide image features  to fine-tune textual features and developed an adaptive textual cache that can be adjusted based on image characteristics, thus achieving instance-wise inference and  enhancing the model's adaptive capabilities.

\item We developed a learnable visual cache that decouples new and prior knowledge, enabling our \itshape ATC \rm to attain maximum acquisition of new knowledge while preserving previously acquired knowledge.

\item Our proposed two-branch structure reduces the over-dependence on LVLMs adopted by existing methods.
\end{itemize}
\begin{figure*}[!t]
\centering
\includegraphics[width=13.6 cm]{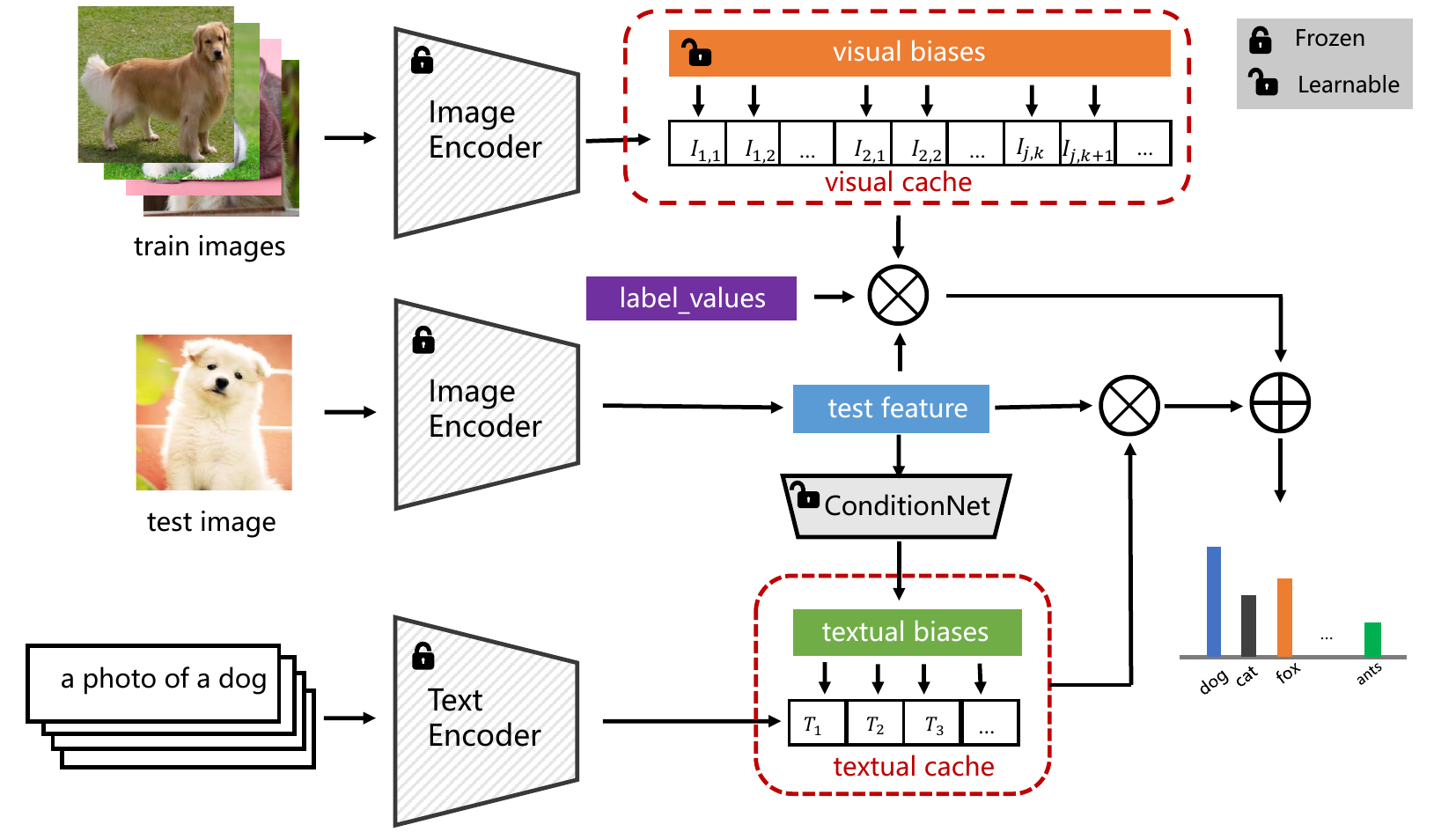}
\caption{\textbf{The Pipeline of \itshape ATC\rm}. Our proposed model adopts a two-branch structure. The upper branch constructs a learnable visual cache by applying an adjustable training matrix to all support image features. The lower branch features our proposed $ConditionNet$ that generates biases for the test image features, which adjusts the textual cache. Combining the two branches results in the production of the final output, thereby addressing the limitations of existing methods that rely only on LVLMs.}
\label{fig 1}
\end{figure*}
\section{RELATED WORK}
\subsection{Large-Scale Vision-Language Models}
Large-scale vision-language models (LVLMs)  combine visual and textual inputs, enabling them to process, understand, and generate associations between images and natural language text, such as CLIP\cite{radford2021learning}, ALBEF\cite{li2021align}, ALIGN\cite{cohen1997align} and BEiT-v3\cite{wang2022image}. These models are pre-trained on large datasets containing text and images to develop an understanding of both types of inputs. We will use CLIP\cite{radford2021learning} as an example to elaborate on.

CLIP\cite{radford2021learning} is one of the most popular LVLMs and is pre-trained on 400 million image-text pairs using contrastive learning techniques\cite{chen2020simple,he2020momentum}. Additionally, it  exhibits strong zero-shot classification ability. The CLIP\cite{radford2021learning} model's training methodology is grounded in two assumptions: (i) text and images can complement each other's information; (ii) different images and texts can be compared for similarity. To achieve this goal, the CLIP\cite{radford2021learning}  uses self-supervised and contrastive learning methods to obtain model parameters by learning the similarity relationship between images and text features. During pretraining, CLIP\cite{radford2021learning} maps images and text to the same space to update the model parameters by comparing the similarity of positive and negative samples. During the inference stage, given an image and a series of image captions, they are respectively processed by the visual encoder and textual encoder to obtain the image feature vector $z$ and the text feature vector $t$. Then, their cosine similarity is calculated using the following formula:
\begin{equation}
p(y = i|z) = \frac{exp(sim(z, t_i)/\tau )}{ \sum ^K _{j=1} {exp(sim(z, t_j)/\tau )}} 
\end{equation}
where  $ sim(\cdot ,\cdot ) $ indicates cosine similarity, and $\tau$ is the learned temperature of CLIP\cite{radford2021learning}. Recently, researchers have found that utilizing text supervision can greatly improve the visual representation ability of models. Our work aims to transfer large-scale pre-trained vision-language models to specific tasks  via text supervision and a small amount of training data.
\subsection{Data-efficient Transfer Learning}
Data-efficient transfer learning is a subfield of machine learning that facilitates prior knowledge from pre-trained  models to be applied to new target tasks while employing limited training data. Given the financial and temporal costs of acquiring and annotating sizable data, data-efficient transfer learning has emerged as a prevalent research domain. In this regard, pretrained  models obtained from large-scale datasets, such as ImageNet\cite{deng2009imagenet}, by utilizing advanced hardware and neural network structures might be used to expedite the target task's learning process. However, because of the discrepancy in data and distributions between source and target tasks, achieving optimal performance becomes a challenge, and several strategies are employed in data-efficient transfer learning, such as  meta-learning and data augmentation. 

Recently, new paradigms have emerged with the increasing number of large-scale vision-language models. CoOp\cite{zhou2022learning} and CoCoOp\cite{zhou2022conditional} proposed to replace the fixed prompt templates in large-scale vision-language models with continuous and learnable vectors, and fine-tune them based on downstream task data. CLIP-Adapter\cite{gao2021clip} uses simple linear projection to adapt features to downstream tasks. TaskRes\cite{yu2022task} improves model performance by decoupling old and new knowledge of large-scale vision-language models through a simple approach. In this work, we propose a new two-branch model for data-efficient  transfer learning that achieves better performance than the aforementioned methods.

\section{METHOD}
 The overall pipeline of our proposed method is shown in  Figure  \ref{fig 1}. In this section, we will first review the challenges of current data-efficient transfer learning based on LVLMs, and then specifically introduce our proposed method, \itshape ATC\rm.
\subsection{Defects of Existing Data-efficient Transfer Learning based on LVLMs}
LVLMs  possess massive parameters, and fine-tuning the model with insufficient data presents a risk of over-fitting and causes damage to model's original knowledge. However, due to domain shift between the training data of LVLMs and downstream tasks, transfer learning is necessary. Thus, effective data-efficient  transfer learning should enable the model to acquire new knowledge substantively while upholding its previous knowledge. However, there are still some issues that need to be addressed when applying LVLM-based  data-efficient transfer learning currently.

Methods that employ  prompt engineering at input level, such as CoOp\cite{zhou2022learning} and CoCoOp\cite{zhou2022conditional}, allow models to alter static text prompt templates into continuous learnable prompt templates for adaptation to new tasks. Nevertheless, using these methods could lead to loss of some of CLIP's original knowledge. For example, CoOp's\cite{zhou2022learning} 1-shot and 2-shot accuracy is inferior to CLIP's\cite{radford2021learning} zero-shot accuracy, while the authors of CoCoOp\cite{zhou2022conditional} reported that the computing resources and time consumed by the model during training are considerable. Even though the visual and textual encoders are set with fixed parameters, and the model only modifies a few parameters, saving the complete model's gradient is resource-intensive during training for backpropagation. Similarly, Hyojin\cite{bahng2022exploring} and MaPLe\cite{khattak2022maple} introduced prompt engineering\cite{liu2021gpt,shin2020autoprompt,jiang2020can,li2021prefix} from natural language processing into image processing, which sacrifices prior knowledge and consumes resources for LVLMs to adapt to downstream tasks.

 Approaches that fine-tune features on the feature side, such as  CLIP-Adapter\cite{gao2021clip} only utilize adapters after the visual or textual encoder, which maps new knowledge onto old knowledge in an excessively coupled manner. Due to this excessive coupling, the learning capacity for new knowledge is limited. Although TaskRes\cite{yu2022task} decouples new and old knowledge by implementing learnable masks on text features, the mask remains static after the completion of training, leading to limited adaptability. Moreover, both paradigms inevitably suffer from a problem where the final classification result only relies on the similarity between visual and textual features, resulting in the inadequate utilization of training data and excessive dependence on LVLMs. Moreover, both paradigms have poor adaptability because text features remain unchanged once calculated and cannot be adjusted based on image characteristics.
\subsection{\itshape ATC \rm}
In response to the issues with existing methods, we propose a novel two-branch model. On one branch, we use training data to construct a learnable visual cache. On the other branch, we employ our proposed $ConditionNet$ to adjust the adaptive textual cache. The final result is jointly determined by these two branches. In the following sections, we will provide a detailed introduction to our \itshape ATC\rm.
\subsubsection{$ConditionNet$}

Even the visual features of images in the same category vary, class-wise text features are insufficient to cover the significant changes in the appearance and context geometry of the current category, making it impossible to adapt to the unique features of each test image. Therefore, if only fixed and invariant class-wise text features are used as the final classification criteria, the model's generalization and stability will be compromised. Therefore, to enable text features to automatically adjust based on the features of test images, we propose $ConditionNet$. This network can  generate textual biases to adjust the textual cache by perceiving the features of test images. The proposed $ConditionNet$ facilitates communication between image features and text features, achieving adaptive adjustment of text cache during the testing process. By utilizing instance-wise inference, it improves the model's generalization and adaptability. Here, we use a small number of parameters  LSTM as the $ConditionNet$.

\subsubsection{Adaptive Textual cache}
We propose an adaptive textual cache that can fine-tune the text features based on the characteristics of test images which disrupts the paradigm of traditional methods, where the text features remain constant once computed. Here is the process of constructing the adaptive textual cache. Firstly, the category labels of the entire dataset are placed in a fixed manual prompt template (e.g., “”a photo of a \{class\}”), denoted by $V$, $V=[v_1,v_2,\dots ,v_c]$, then, $V$ is fed into the textual encoder of the CLIP\cite{radford2021learning}, $Encoder\_{txt}$, to obtain the initial textual cache $P_{txt}=[t_1,t_2,\dots,t_c ]\subset R^{c\times dim}$. Here, $c$ represents the number of categories in the dataset, and $dim$ represents the output dimension of the $Encoder\_{txt}$. Calculating the cache only once significantly reduces computational cost. Meanwhile, test image $I$ is processed by the visual encoder, $Encoder\_{img}$, resulting in image features, $f_{test} \subset R^{1\times dim}$, the feature $f_{test}$ is fed to our proposed  $ConditionNet$, a model with few parameters, to generate  textual biases, $textual\_biases \subset R^{c\times dim}$,
\begin{equation}
    s=ConditionNet(f_{test})
\end{equation}
\begin{equation}
    textual\_biases=[s,s,\dots,s]
\end{equation}
Overlaying the $textual\_biases$ onto the textual cache generates the latest cache, $\hat{P}_{txt}\subset R^{c\times dim}$,
\begin{equation}
    \hat{P}_{txt}=P_{txt}+textual\_biases
\end{equation}
\subsubsection{Learnable Visual cache}
To eliminate the excessive reliance on LVLMs, we proposed a novel branch, and constructed a learnable visual cache on this branch. Through this cache, our ATC has achieved the decoupling of new and existing knowledge, enabling the model to fully acquire new knowledge while retaining prior knowledge. Without loss of generality, let us consider an experiment with $k$  shots and $ n $ classes, with $ n\times k $ training images that have undergone data augmentation, represented as  $S_{aug}$. We obtain the initial visual cache by encoding the images once, $P_{img}\subset R^{n \cdot k\times dim }$,
\begin{equation}
    P_{img}=Encoder\_img(S_{aug})=[I_{1,1},I_{1,2},\dots I_{i,j} ,\dots I_{c,k}]
\end{equation}
and generate a one-hot matrix of labels for all training data in accordance with the order of the images, denoted as $label\_values \subset R^{n \cdot k \times c}$.  $I_{i,j}$ represents the visual features of the $j-th$ image corresponding to the $i-th$ category. To enable the image cache to be automatically adjusted for the task, we initialize a matrix of zeros, denoted as $visual\_biases\subset R^{n \cdot k \times dim}$, which is automatically updated during model training and aggregated with the original visual cache. 
\begin{equation}
    \hat{P}_{img}=P_{img}+visual\_biases
\end{equation}
We incorporate the design concept of the Tip-Adapter\cite{zhang2022tip} caching model when creating the visual cache. Our method differs from Tip-Adapter\cite{zhang2022tip} in that Tip-Adapter\cite{zhang2022tip} initializes a linear layer with training data, while we overlay a learnable mask initialized at zero on image features from the training data. Our method offers a larger learning space for the model.

~\\
Given an image feature $f_{test}\subset R^{1 \times dim}$, the final predicted probability distribution $f \subset R^{1 \times c}$ is obtained through the above three modules using the following formula:
\begin{equation}
     f_1=f_{test}\otimes  \hat{P}^{\mathsf{T}}_{img}\otimes  label\_values
\end{equation}
\begin{equation}
   f_2=f_{test}\otimes  \hat{P}_{txt}^{\mathsf{T}}
\end{equation}
\begin{equation}
    f=\alpha f_1+\beta f_2
    \label{f 1}
\end{equation}
Here, $\otimes$ represents the hadamard product of matrices, $f_1$ represents the cosine similarity between the $f_{test}$ and the visual cache, while $f_2$ represents the cosine similarity between the $f_{test}$ and the textual feature  cache. The variables $\alpha $ and $\beta $ are weighting coefficients.

\subsection{EXPERIMENT}
\subsection{Experiment Setup}
We follow previous  work\cite{yu2022task,gao2021clip,zhou2022learning} to conduct a few-shot evaluation on 11 benchmark datasets, including ImageNet\cite{deng2009imagenet}, Caltech101\cite{fei2004learning}, OxfordPets\cite{parkhi2012cats}, StanfordCars\cite{krause20133d}, Flowers102\cite{nilsback2008automated}, Food101\cite{bossard2014food}, FGVCAircraft\cite{maji2013fine}, SUN397\cite{xiao2010sun}, DTD\cite{cimpoi2014describing}, EuroSAT\cite{helber2019eurosat}, and UCF101\cite{soomro2012ucf101}. These datasets cover various computer vision tasks, specifically, ImageNet and Caltech101 are used for classification of generic objects, while OxfordPets, StanfordCars, Flowers102, Food101, and FGVCAircraft are used for fine-grained classification, SUN397 is used for scene recognition, UCF101 is used for action recognition, DTD is used for texture classification, and finally, EuroSAT is used for satellite imagery recognition. We train the model by randomly sampling 1/2/4/8/16  samples from each class in the training data and test the model on the entire test dataset. During the training process, the following cross entropy loss is utilized:
\begin{equation}
    L=-[ylog(\hat{y})+(1-y)log(1-\hat{y})]
\end{equation}

Additionally, we performed domain generalization experiments following the experimental settings of CoCoOp\cite{zhou2022conditional}, Tip-Adapter\cite{zhang2022tip}, and TaskRes\cite{yu2022task}. For the domain generalization experiments, we used ImageNet\cite{deng2009imagenet} as the source dataset and four  datasets  ImageNet-V2\cite{recht2019imagenet}, ImageNet-Sketch\cite{wang2019learning}, ImageNet-A\cite{hendrycks2021natural}, and ImageNet-R\cite{hendrycks2021many}  , which have certain domain differences from ImageNet, as our target datasets. 

\begin{figure*}[!t]
	\centering
	\begin{minipage}{0.32\linewidth}
		\centering
		\includegraphics[width=6.4cm]{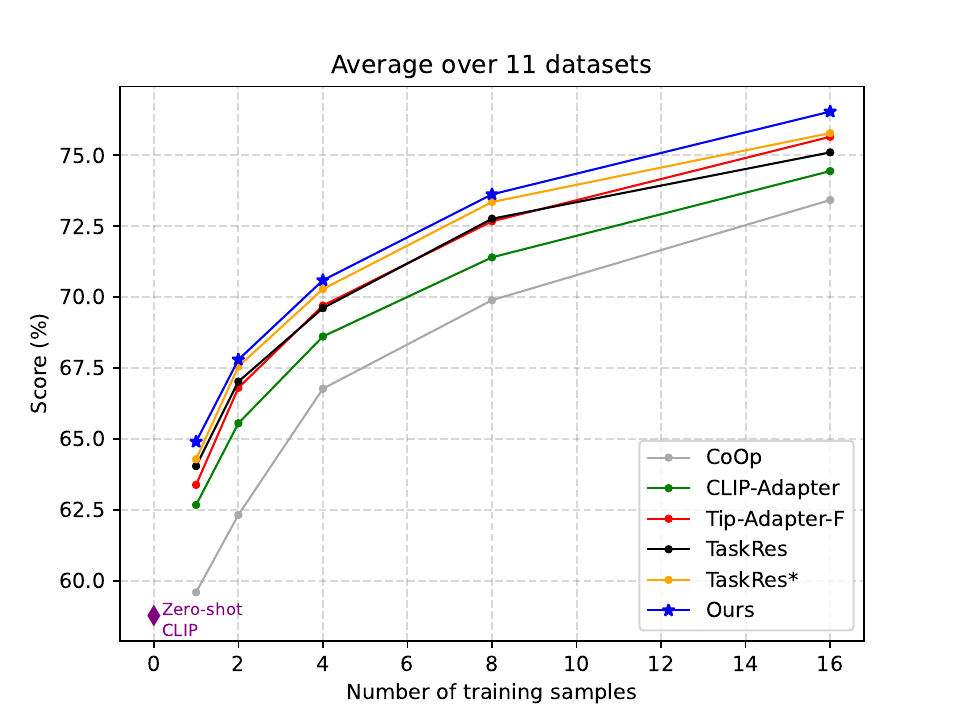}
	\end{minipage}
	\begin{minipage}{0.32\linewidth}
		\centering
		\includegraphics[width=6.4cm]{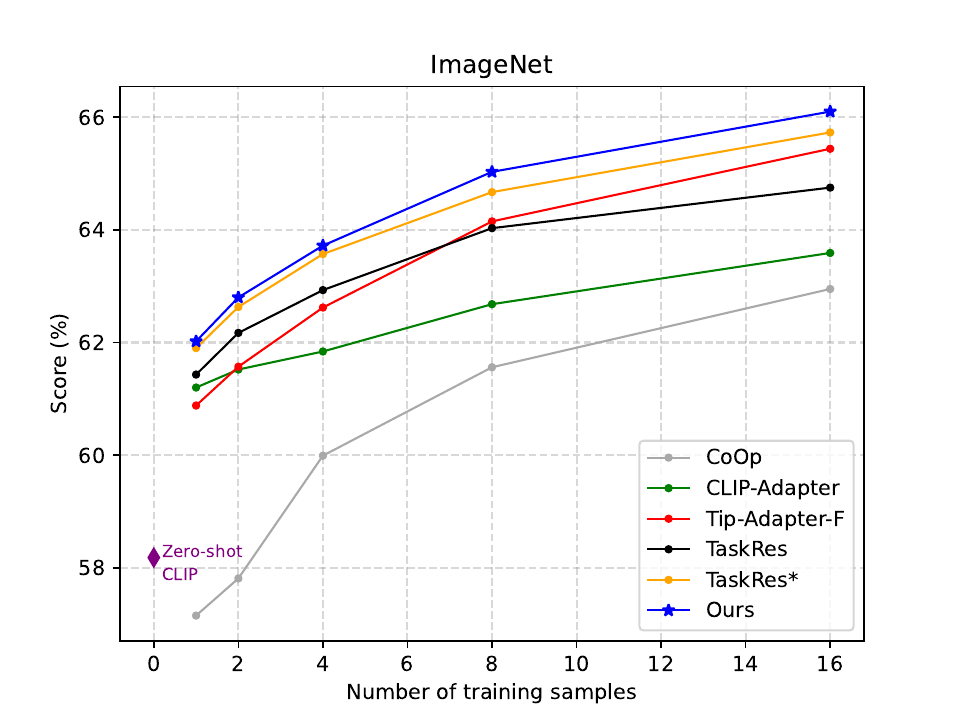}
	\end{minipage}
	\begin{minipage}{0.32\linewidth}
		\centering
		\includegraphics[width=6.4cm]{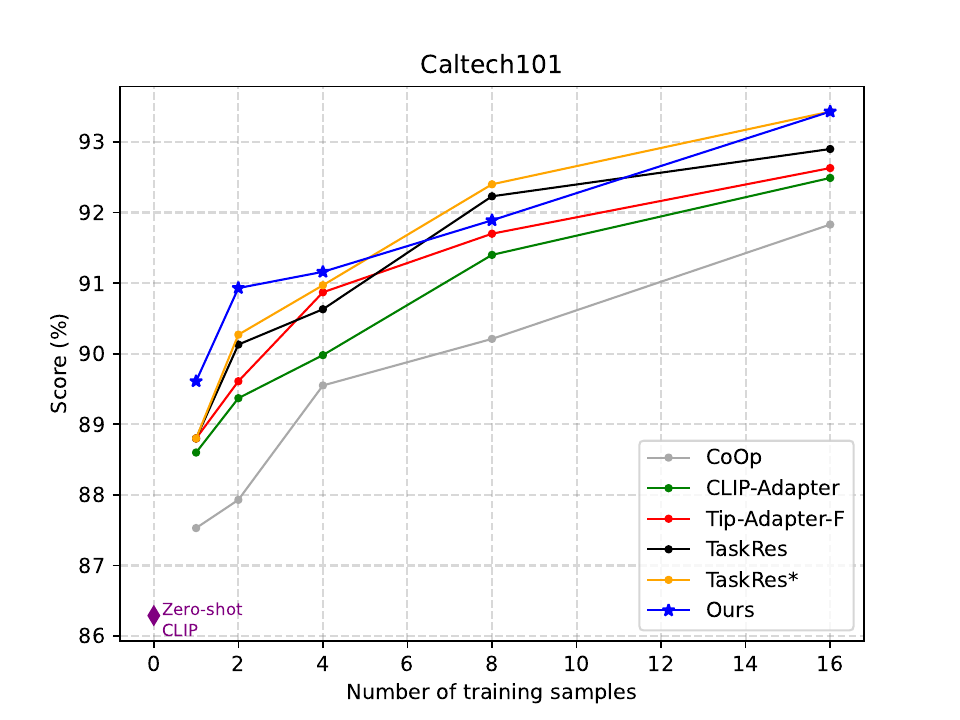}
		
	\end{minipage}
	
	\begin{minipage}{0.32\linewidth}
		\centering
		\includegraphics[width=6.4cm]{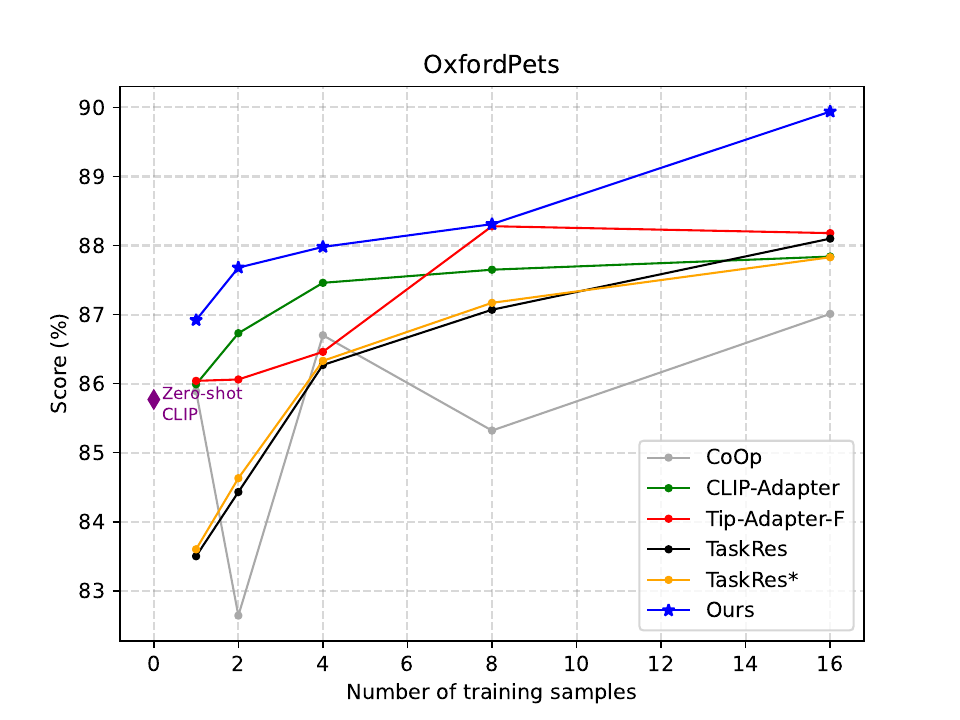}
	\end{minipage}
	\begin{minipage}{0.32\linewidth}
		\centering
		\includegraphics[width=6.4cm]{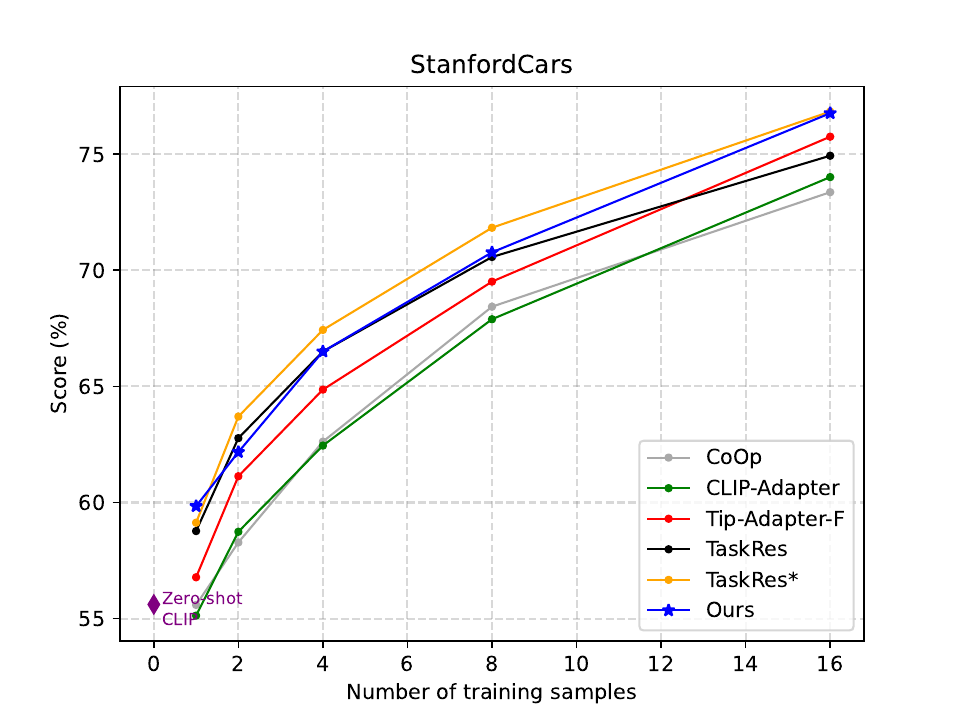}
	\end{minipage}
    \begin{minipage}{0.32\linewidth}
		\centering
		\includegraphics[width=6.4cm]{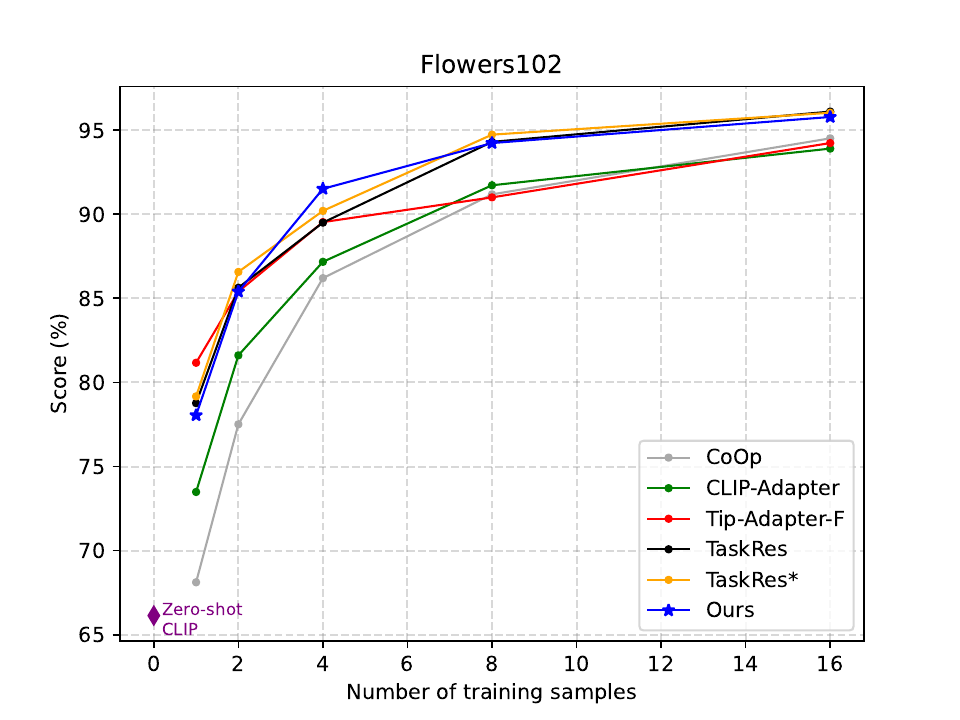}
  
	\end{minipage}
    \begin{minipage}{0.32\linewidth}
		\centering
		\includegraphics[width=6.4cm]{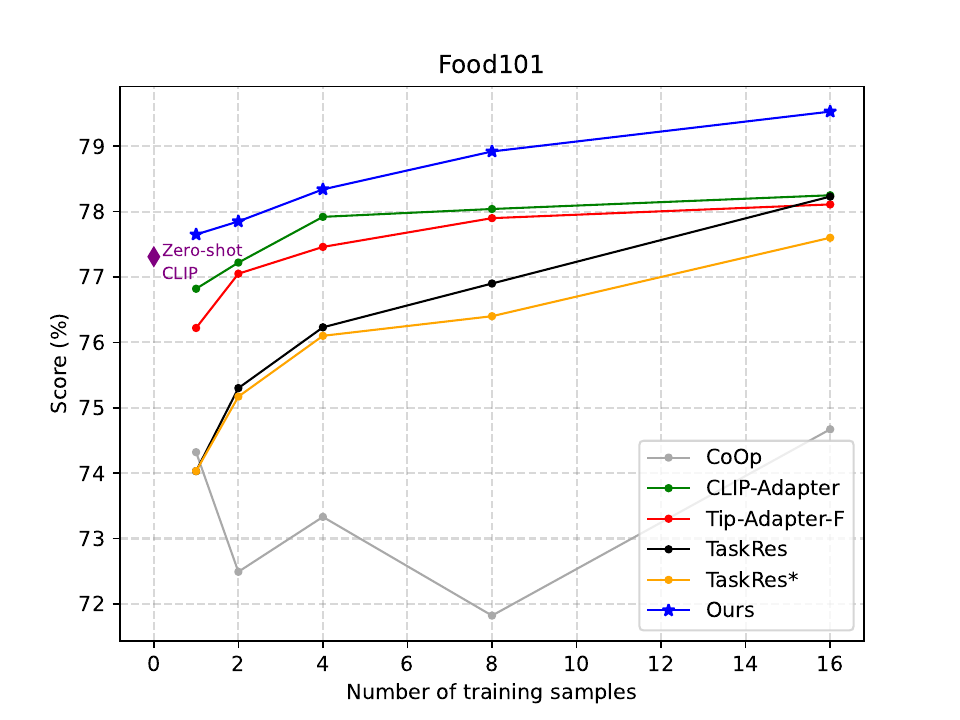}
	\end{minipage}
    \begin{minipage}{0.32\linewidth}
		\centering
		\includegraphics[width=6.4cm]{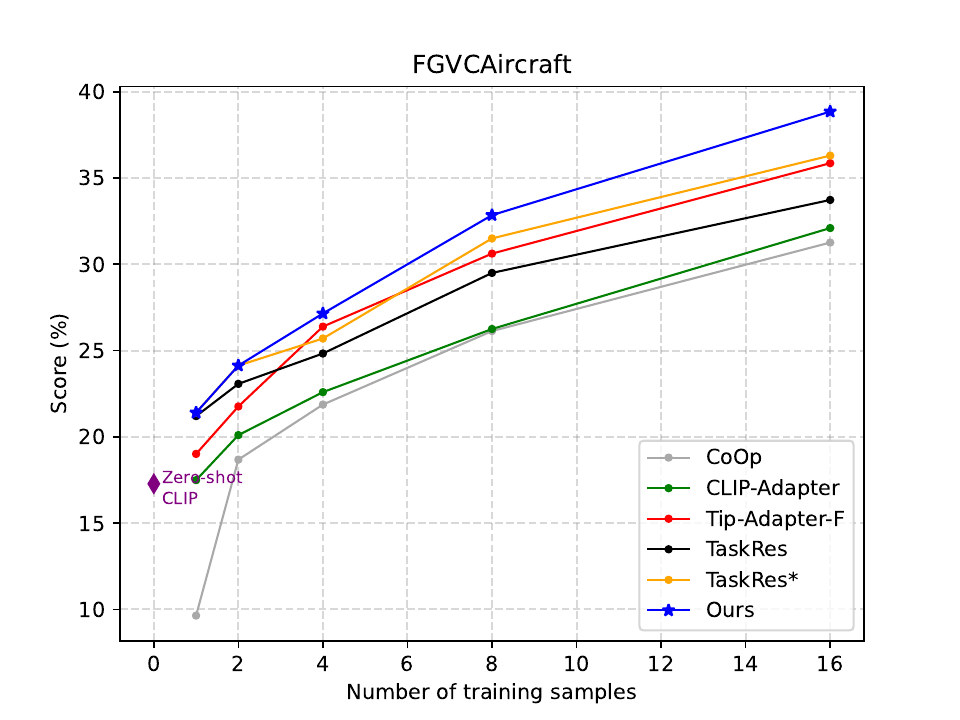}
	\end{minipage}
    \begin{minipage}{0.32\linewidth}
		\centering
		\includegraphics[width=6.4cm]{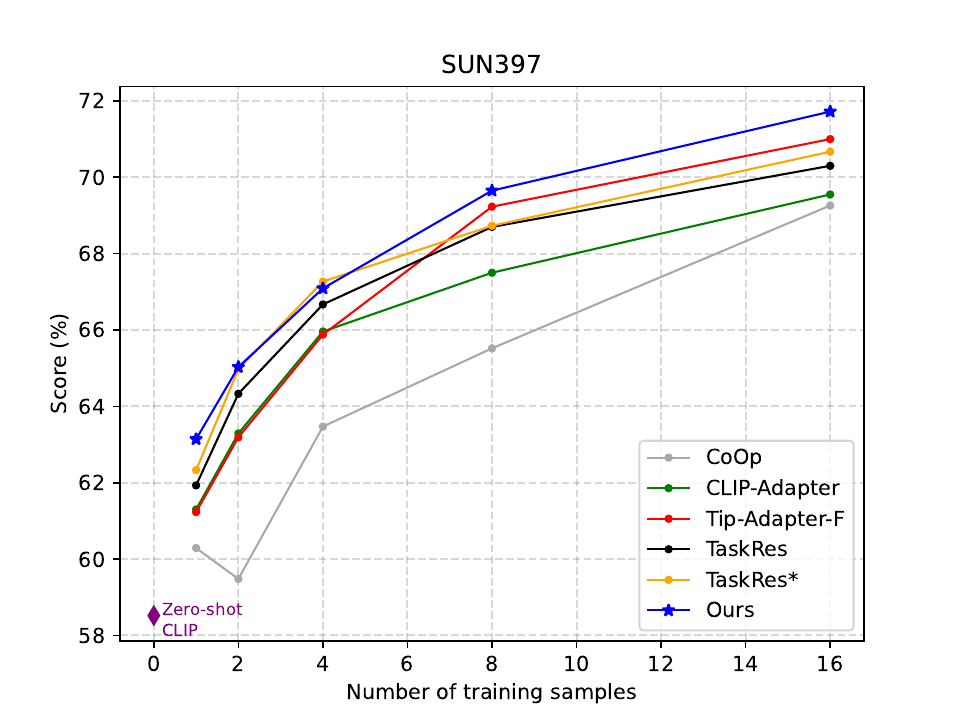}
  
	\end{minipage}
    \begin{minipage}{0.32\linewidth}
		\centering
		\includegraphics[width=6.4cm]{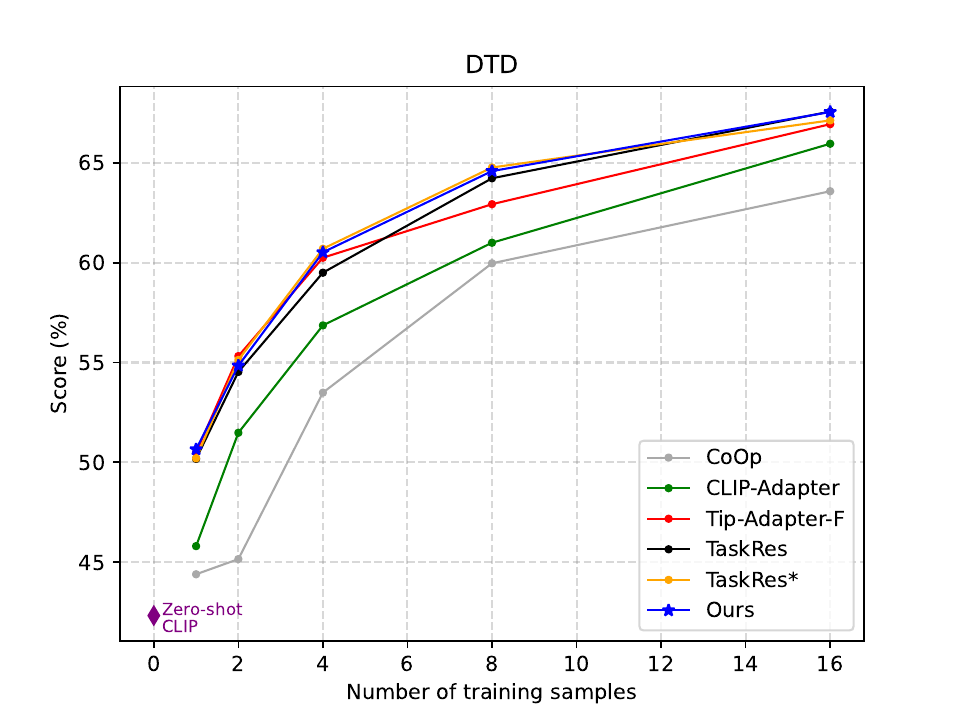}
	\end{minipage}
    \begin{minipage}{0.32\linewidth}
		\centering
		\includegraphics[width=6.4cm]{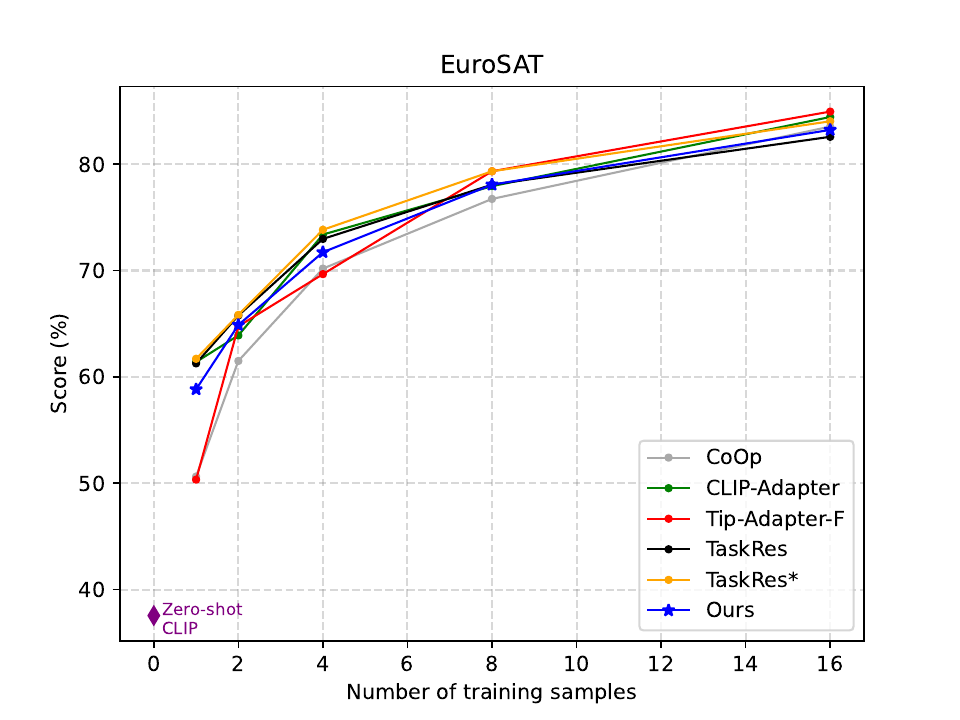}
	\end{minipage}
    \begin{minipage}{0.32\linewidth}
		\centering
		\includegraphics[width=6.4cm]{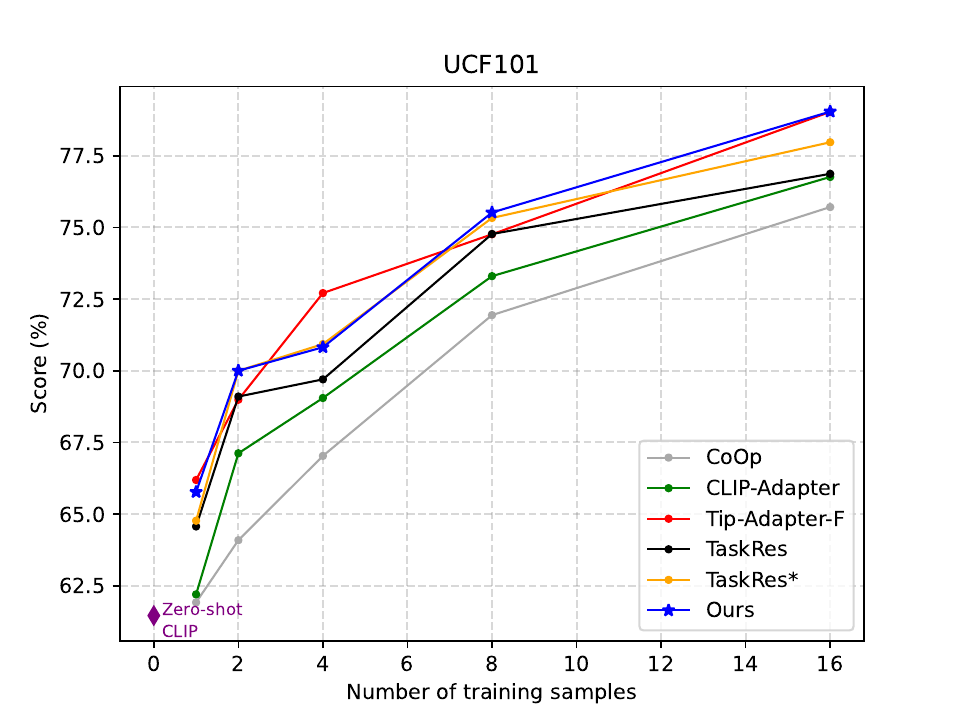}
	\end{minipage}
  
    \caption{\textbf{Main results of few-shot learning on 11 datasets}. Our approach \itshape ATC \rm consistently shows better performance over previous baselines across different training shots, the top-left is the averaged accuracy over the 11 datasets.}
\label{few-shot compare}
\end{figure*}

\begin{table*}[!t]
\centering
\caption{\textbf{Performance comparison on generalization (from ImageNet to ImageNet-V2/-Sketch/-A/-R) with multiple CLIP visual backbones}. Our proposed method achieved the highest average accuracy across four different visual backbones.}
\label{table 1}
\scalebox{1.1}{
\begin{tabular}{lccccccc}
\hline
                        &                 & Source                       & \multicolumn{5}{c}{Target}                                                                                                                               \\ \cline{3-8} 
Method                  & Vision backbone & ImageNet                     & -V2                           & -sketch                       & -A                            & -R                            & Average                      \\ \hline
Zero-shot CLIP\cite{radford2021learning} &                 & 58.18                        & 51.34                        & 33.32                        & 21.65                        & 56.00                        & 40.58                        \\
Linear-Probe CLIP\cite{radford2021learning}       &                 & 55.87                        & 45.97                        & 19.07                        & 12.74                        & 34.86                        & 28.16                        \\
CLIP+CoOp(M=16) \cite{zhou2022learning}        &                 & 62.95                        & 55.11                        & 32.74                        & 22.12                        & 54.96                        & 41.23                        \\
CLIP+CoOp(M=4)\cite{zhou2022learning}          & ResNet-50\cite{he2016deep}       & 63.33                        & 55.40                        & 34.67                        &  \textbf{23.06}& 56.60                        & 42.43                        \\
TaskRes*\cite{yu2022task}                &                 & 65.73                        & \textbf{57.00 }& 34.43                        & 21.50                        & 58.13                        & 42.77                        \\
\rowcolor[HTML]{E7E6E6} 
ours                    &                 &  \textbf{66.10} & 56.78                        & \textbf{35.39} & 22.19                        &\textbf{59.25} &  \textbf{43.40} \\ \hline
\rowcolor[HTML]{FFFFFF} 
Zero-shot CLIP\cite{radford2021learning}          &                 & 61.62                        & 54.81                        & 38.71                        & 28.05                        & 64.38                        & 46.49                        \\
Linear-Probe CLIP\cite{radford2021learning}       &                 & 59.75                        & 50.05                        & 26.80                        & 19.44                        & 47.19                        & 35.87                        \\
CLIP+CoOp(M=16) \cite{zhou2022learning}       &                 & 66.60                        & 58.66                        & 39.08                        & 28.89                        & 63.00                        & 47.41                        \\
CLIP+CoOp(M=4)\cite{zhou2022learning}          & ResNet-101\cite{he2016deep}       & 65.98                        & 56.80                        & 40.40                        & 29.60                        & 64.98                        & 47.95                        \\
TaskRes*\cite{yu2022task}                &                 & 68.73                        & 60.00                        & 40.30                        & 28.00                        & 64.80                        & 48.28                        \\
\rowcolor[HTML]{E7E6E6} 
ours                    &                 &  \textbf{69.28}  & \textbf{61.20}&  \textbf{41.55} &  \textbf{30.51} &  \textbf{67.81} &\textbf{50.27 }\\ \hline
\rowcolor[HTML]{FFFFFF} 
Zero-shot CLIP\cite{radford2021learning}          &                 & 62.25                        & 54.79                        & 40.82                        & 29.57                        & 65.99                        & 47.79                        \\
\rowcolor[HTML]{FFFFFF} 
Linear-Probe CLIP\cite{radford2021learning}       &                 & 59.58                        & 49.73                        & 28.06                        & 19.67                        & 47.20                        & 36.17                        \\
CLIP+CoOp(M=16)\cite{zhou2022learning}         &                 & 66.85                        & 58.08                        & 40.44                        & 30.62                        & 64.45                        & 48.40                        \\
CLIP+CoOp(M=4)\cite{zhou2022learning}          & ViT- B/32\cite{dosovitskiy2020image}       & 66.34                        & 58.24                        & 41.48                        & 31.34                        & 65.78                        & 49.21                        \\
TaskRes*\cite{yu2022task}                &                 & 69.17                        & 59.47                        & 40.87                        & 29.70                        & 66.27                        & 49.08                        \\
\rowcolor[HTML]{E7E6E6} 
ours                    &                 &  \textbf{69.45}&  \textbf{61.09} & \textbf{42.10} & \textbf{32.13}                        & \textbf{68.82}                        & \textbf{51.04}                        \\ \hline
Zero-shot CLIP\cite{radford2021learning}          &                 & 66.73                        & 60.83                        & 46.15                        & 47.77                        & 73.96                        & 57.18                        \\
Linear-Probe CLIP\cite{radford2021learning}       &                 & 65.85                        & 56.26                        & 34.77                        & 35.68                        & 58.43                        & 46.29                        \\
\rowcolor[HTML]{FFFFFF} 
CLIP+CoOp(M=16)\cite{zhou2022learning}         &                 & 71.92                        & 64.18                        & 46.71                        & 48.41                        & 74.32                        & 58.41                        \\
CLIP+CoOp(M=4)\cite{zhou2022learning}          & ViT- B/16\cite{dosovitskiy2020image}       & 71.73                        & 64.56                        & 47.89                        & 49.93                        & 75.14                        & 59.38                        \\
TaskRes* \cite{yu2022task}               &                 & 73.90                        & 65.85                        & 47.70                        & 49.17                        & 75.23                        & 59.49                        \\
\rowcolor[HTML]{E7E6E6} 
ours                 &                 &\textbf{ 74.34  }                      & \textbf{ 66.02 }                       & \textbf{ 48.89 }                       & \textbf{ 50.38}                        & \textbf{ 77.38 }                       & \textbf{ 60.67  }       
\end{tabular}
}

\end{table*}

\subsection{Baseline models}
For the few-shot classification experiment, we compare our approach with Zero-shot CLIP\cite{radford2021learning}, Linear-Probe CLIP\cite{radford2021learning}, CoOp\cite{zhou2022learning}, Tip-Adapter-F\cite{zhang2022tip}, TaskRes\cite{yu2022task}, and TaskRes*\cite{yu2022task}. Both Zero-shot CLIP and Linear-Probe CLIP\cite{radford2021learning} utilized the same handcrafted prompt template, such as "a photo of a \{class\}", while CoOp\cite{zhou2022learning} replaced the fixed handcrafted prompt template with a continuous learnable template.  Tip-Adapter-F\cite{zhang2022tip} represents a variant of Tip-Adapter\cite{zhang2022tip} that utilizes a small amount of data to train a caching model. TaskRes\cite{yu2022task} improves the classification performance of models by augmenting the text features with a learnable matrix initialized at zero, while TaskRes*\cite{yu2022task} further boosts the performance by enhancing the classifier based on TaskRes\cite{yu2022task}.

For the domain generalization experiment, we compare our approach with Zero-shot CLIP\cite{radford2021learning}, Linear-Probe CLIP\cite{radford2021learning}, CLIP+CoOp (M=16)\cite{zhou2022learning}, CLIP+CoOp (M=4)\cite{zhou2022learning}, TaskRes*\cite{yu2022task}. 
\subsection{Performance Comparison}
\begin{figure}[H]
\centering
\includegraphics[width=8cm]{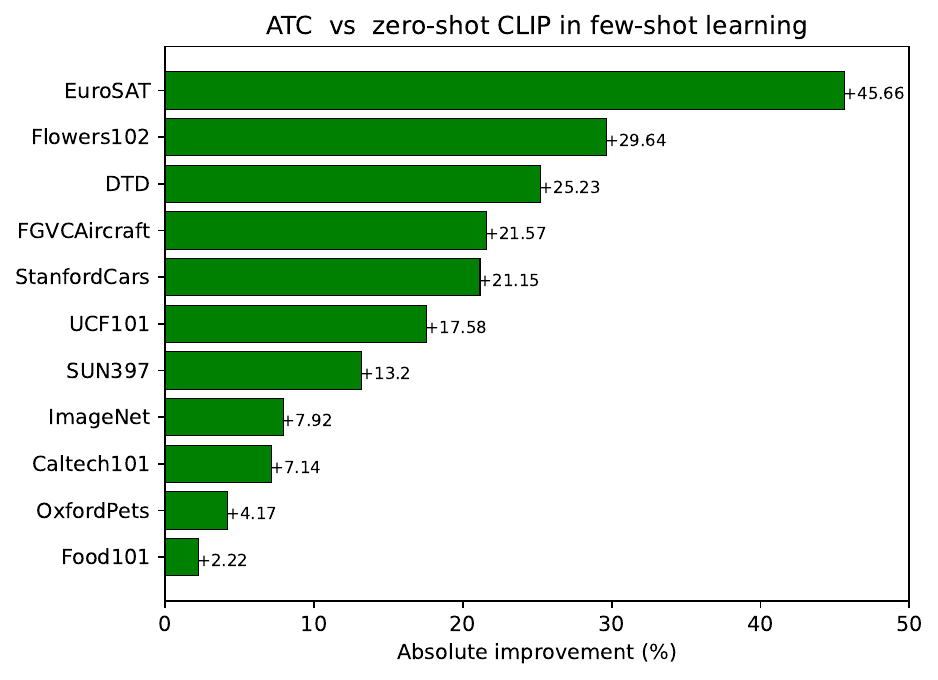}
 
\caption{\textbf{Comparison with Zero-shot CLIP\cite{radford2021learning} in few-shot setting}. On 11 benchmark datasets, our method demonstrates a significant improvement in accuracy compared to Zero-shot CLIP.}
\label{fig 3}
\end{figure}

\paragraph{few-shot learning}
The primary experimental results, as depicted in Figure \ref{few-shot compare}, showcase a comparison between our proposed 1/2/4/8/16-shot experimental outcomes and the state-of-the-art few-shot transfer learning techniques based on CLIP\cite{radford2021learning}, including Linear-Probe CLIP\cite{radford2021learning}, CoOp\cite{zhou2022learning}, Tip-Adapter-F\cite{zhang2022tip}, TaskRes\cite{yu2022task}, and TaskRes*\cite{yu2022task}. Our method achieves the highest average classification accuracy on 11 datasets, as indicated in the upper left corner of the Figure \ref{few-shot compare} , and notably outperforms other algorithms on prominent datasets such as ImageNet, OxfordPets, Food101, FGVCAircraft, SUN397, especially on Food101, where other methods fail to demonstrate satisfactory results. Remarkably, our approach achieves the highest accuracy on eight datasets out of eleven, excluding StanfordCars, Flowers102, and EuroSAT, upon conducting 16-shot experiments. Furthermore, our proposed method outperforms Zero-shot CLIP\cite{radford2021learning} on all 11 datasets, as depicted in Figure \ref{fig 3}, as compared to the Zero-shot CLIP\cite{radford2021learning}, our approach has significant improvement in  accuracy.

\subsubsection{Domain Generalization}

\begin{figure}[!t]
\centering
\includegraphics[width=8cm]{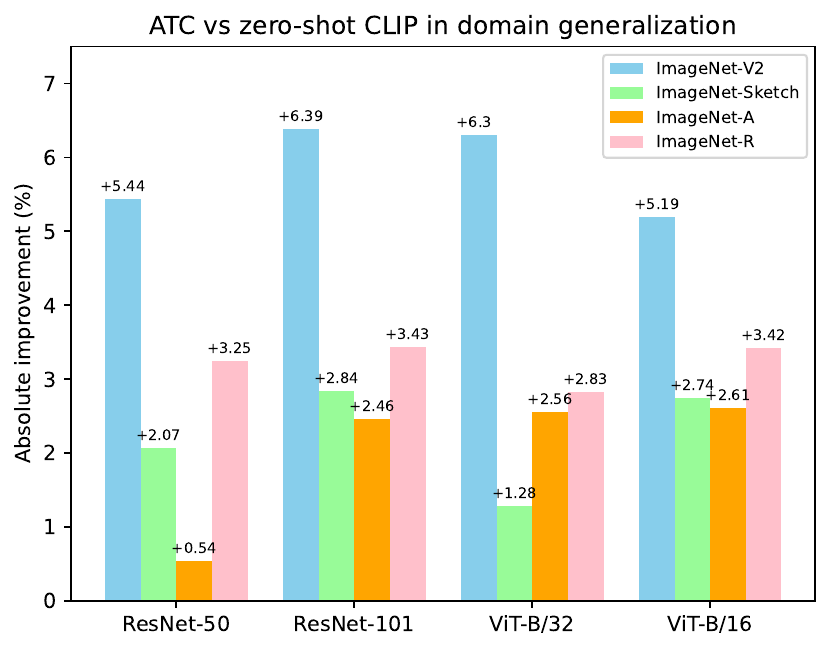}

\caption{\textbf{Comparisions with Zero-shot CLIP\cite{radford2021learning} in domain generalization}. Our  \itshape ATC \rm  has achieved improvement in accuracy compared to zero-shot CLIP on four benchmark datasets by using different visual backbone networks.}

\vspace{-0.2cm}

\label{fig 4}
\end{figure}

The primary objective of this experiment is to assess the model's generalization ability. We randomly select 16 samples per class from ImageNet as training data and independently test our model on the entire test sets of variants ImageNet datasets, including ImageNet-V2, ImageNet-Sketch, ImageNet-A, and ImageNet-R. We compare our experimental results with those of other state-of-the-art few-shot transfer learning techniques, including Zero-shot CLIP\cite{radford2021learning}, Linear-Probe CLIP\cite{radford2021learning}, CLIP+CoOp(M=16)\cite{zhou2022learning}, CLIP+CoOp(M=4)\cite{zhou2022learning}, and TaskRes*\cite{yu2022task}. Our method achieves the highest average accuracy on all four datasets by using different visual backbone networks except for ResNet-50\cite{he2016deep}, as shown in the summarized results in Table \ref{table 1}. When used as the visual encoder, ResNet-50's\cite{he2016deep} encoding dimension is 1024, while other visual backbone networks have an encoding dimension of only 512, which results in unsatisfactory performance. Therefore, using ResNet-50\cite{he2016deep} as the visual backbone network requires the model to learn more parameters when constructing the adaptive textual cache and $ConditionNet$, leading to some degree of over-fitting on the source dataset. Furthermore, as illustrated in Figure \ref{fig 4}, our approach demonstrated excellent generalizability as evidenced by its higher accuracy on all four target datasets compared to Zero-shot CLIP\cite{radford2021learning}.

\subsubsection{Ablation Study}

In this section, we mainly conducted ablation experiments on the two main modules we proposed  adaptive textual cache and learnable visual cache and four different visual backbone networks, namely ResNet-50\cite{he2016deep}, ResNet-101\cite{he2016deep}, ViT-B/32\cite{dosovitskiy2020image}, and ViT-B/16\cite{dosovitskiy2020image}, using different weighting coefficients $\alpha$ and $\beta$. The experiments were conducted primarily on the ImageNet dataset.
\paragraph{Adaptive Textual cache}
We conducted ablation experiments on adaptive textual cache by comparing it with fixed textual cache. Table \ref{table 2} summarizes the experimental results, indicating that using adaptive textual cache improves accuracy across 1/2/4/8/16-shot experiments, effectively validating the efficacy of this module.
\begin{table}
\centering
\caption{\textbf{Ablation Study of adaptive textual cache}.In different few-shot experimental settings, utilizing an adaptive textual cache is beneficial in enhancing the model's performance.}
\begin{tabular}{lccccc}
\hline
Textual cache & 1-shot & 2-shot & 4-shot & 8-shot & 16-shot \\ \hline
fixed               & 61.79  & 62.45  & 63.23  & 64.42  & 65.72   \\
adaptive            &\textbf{62.02}  & \textbf{62.80}   & \textbf{63.72}  &\textbf{65.03}  & \textbf{66.10}    \\ \hline
\textbf{}           & \textcolor{blue}{+0.23}  & \textcolor{blue}{+0.35}  & \textcolor{blue}{+0.49}  & \textcolor{blue}{+0.61}  & \textcolor{blue}{+0.38}   \\ \hline

\end{tabular}

\label{table 2}
\end{table}
\paragraph{Learnable Visual cache}

The effectiveness of the learnable visual cache is validated through experiments. We conducted ablation experiments on the ImageNet dataset to compare our method of constructing visual caches by adding learnable biases with the methods of constructing fixed caches and using learnable linear layers in Tip-Adapter-F\cite{zhang2022tip}, the experimental results are presented in the Table \ref{table 3}. The experimental results are as follows: from the experimental results, it can be seen that our proposed method of constructing caches outperforms the linear layer initialization in Tip-Adapter-F, demonstrating the effectiveness of our proposed method.

\begin{table}[]
\caption{\textbf{Ablation study of learnable visual cache}. Under various few-shot experimental settings, constructing a learnable visual cache by adding biases can greatly enhance the model's performance.}
\begin{tabular}{lccccc}
\hline
Visual   cache         & 1-shot         & 2-shot         & 4-shot         & 8-shot         & 16-shot        \\ \hline
fixed                  & 61.87          & 62.39          & 62.78          & 63.75          & 64.68          \\
learnable   linear layer & 61.96          & 62.31          & 62.78          & 64.49          & 65.84          \\
adding  biases         & \textbf{62.02}          & \textbf{62.80}          & \textbf{63.72}          & \textbf{65.03}         & \textbf{66.10}        \\ \hline
\textbf{}              & \textcolor{blue}{+0.06} & \textcolor{blue}{+0.49} & \textcolor{blue}{+0.94} & \textcolor{blue}{+0.54} & \textcolor{blue}{+0.26} \\ \hline
\end{tabular}

\label{table 3}
\end{table}

\paragraph{Vision Backbone}
Furthermore, we conducted few-shot experiments on ImageNet dataset for different visual backbone networks, including ResNet-50\cite{he2016deep}, ResNet-101\cite{he2016deep}, ViT-B/32\cite{dosovitskiy2020image} and ViT-B/16\cite{dosovitskiy2020image}. The experimental results are presented in the Table \ref{table 4}, which demonstrates that our method outperforms other methods across all visual backbone networks.
\begin{table}[!h]
\caption{Results of CLIP visual backbones on 16-shot ImageNet.}
\centering
\begin{tabular}{lcccc}
\hline
Method         & ResNet-50& ResNet-101 & ViT-B/32 & ViT-B/16\\ \hline
Zero-shot CLIP\cite{radford2021learning} & 58.18     & 61.62      & 62.05    & 66.73    \\
CoOp\cite{zhou2022learning}           & 62.95     & 66.60      & 66.85    & 71.92    \\
CLIP-Adapter\cite{gao2021clip}   & 63.59     & 65.39      & 66.19    & 71.13    \\
Tip-Adapter-F\cite{zhang2022tip}  & 65.44     & 68.56      & 68.65    & 73.69    \\
TaskRes\cite{yu2022task}        & 64.75     & 67.70      & 68.20    & 73.07    \\
TaskRes*\cite{yu2022task}       & 65.73     & 68.73      & 69.17    & 73.90    \\
\rowcolor[HTML]{E7E6E6} 
ours       & \textbf{66.10}    & \textbf{69.10}      & \textbf{69.65}    &\textbf{74.34}    \\ \hline
               & \textcolor{blue}{+0.37}     & \textcolor{blue}{+0.37}      & \textcolor{blue}{+0.48}    & \textcolor{blue}{+0.44}    \\ \hline
\end{tabular}

\label{table 4}
\end{table}

\paragraph{$\alpha$   and   $\beta$}
Based on the Figure \ref{fig 1} and Formula \ref{f 1}, the final classification probability is obtained by combining two branches, with $\alpha$ and $\beta$ as the corresponding weighting coefficients. We conducted separate experiments on these coefficients. When we varied $\alpha$, we fixed $\beta$  at \textbf{1}, and when we varied $\beta$, we fixed $\alpha$ at \textbf{1}. According to the experimental results, the highest accuracy was achieved when both coefficients were set to \textbf{1}. To validate our findings, we conducted 16-shot experiments using ResNet-50\cite{he2016deep} on the ImageNet dataset, the result of experiment are present in Table \ref{table 5}.

\begin{table}
\caption{\textbf{Ablation study of $\alpha$  and $\beta$}. The best performance is achieved when $\alpha$  and $\beta$ are both equal to 1.}
\centering
\begin{tabular}{cccccc}
\hline
$\alpha$        & 0.00  & 0.50  & 1.00  & 1.50  & 2.00  \\
accuracy & 58.18 & 65.03 & \textbf{66.10} & 65.89 & 65.98 \\ \hline
$\beta$        & 1.00  & 1.50  & 2.00  & 2.50  & 3.00  \\
accuracy & \textbf{66.10} & 66.01 & 65.92 & 65.75 & 64.26\\ \hline
\end{tabular}

\label{table 5}

\end{table}

\section{Conclusion, Limitations and Future Work}
Our proposed method aims to address the primary issue of data-efficient transfer learning found in LVLMs. We propose a new two-branch model called ATC. In the first branch, We constructed a learnable visual cache from the training data, which enables the decoupling of new and old knowledge. This allows our model to acquire new knowledge while retaining the prior knowledge of LVLMs. In the second branch, the proposed $ConditionNet$ guides visual features to generate textual biases that are overlaid on the text feature, creating an adaptive textual cache. This cache automatically adjusts the text feature based on the image feature, providing strong adaptability. The final output of the model combines the two branches, reducing the existing methods' over-reliance on LVLMs, and our approach generates higher accuracy on 11 datasets  compared to previous methods.

However, our proposed method has one primary limitation, an increase in training data will result in an visual cache expansion, leading to a higher number of parameters and resource consumption.

The integration of vision-language, and multimodal pre-training is a growing field and requires further research and exploration to efficiently transfer various large-scale models to downstream tasks. We anticipate that the empirical findings and insights we present here can lay the groundwork for future research on efficient adaptation methods for emerging fundamental models, which still require more investigation.

\ack This work was supported in part by the National Key R\&D Program of China under Grant 2021ZD0112000, the National Natural Science Foundation of China under Grant 62271119, the Natural Science Foundation of Sichuan Province under Grant 2023NSFSC1972.

\bibliography{ecai}
\end{document}